\documentclass[sigconf]{acmart}

\def\BibTeX{{\rm B\kern-.05em{\sc i\kern-.025em b}\kern-.08emT\kern-.1667em\lower.7ex\hbox{E}\kern-.125emX}}
    
\copyrightyear{2019}
\acmYear{2019}
\setcopyright{acmlicensed}
\acmConference[COINS 2019]{International Conference on Omni-Layer Intelligent Systems}{May 05--07, 2019}{Crete, GREECE}

\begin{document}

%
\title{Blockchain-based Smart-IoT Trust Zone Measurement Architecture}

%
\author{Jawad Ali}
\authornote{Corresponding author.}
\affiliation{%
  \institution{Malaysian Institute of Information Technology, \\ Universiti Kuala Lumpur}
  \city{Kuala Lumpur}
  \country{Malaysia}}
\email{jawad.ali@s.unikl.edu.my}

\author{Toqeer Ali \&}
\author{Yazed Alsaawy}
\affiliation{%
  \institution{Faculty of Computer \& Information Systems, \\ Islamic University of Madinah}
  \streetaddress{Address}
  \city{Madinah}
  \country{Saudi Arabia}}
\email{toqeer@iu.edu.sa}
\email{yalsaawy@iu.edu.sa}


\author{Ahmad Shahrafidz Khalid \&}
\author{Shahrulniza Musa}
\affiliation{%
	\institution{Malaysian Institute of Information Technology, \\ Universiti Kuala Lumpur}
	\streetaddress{Address}
	\city{Kuala Lumpur}
	\country{Malaysia}}
\email{ahmads@unikl.edu.my}
\email{shahrulniza@unikl.edu.my}

\renewcommand{\shortauthors}{J.Ali, et al.}
\copyrightyear{2019}
\acmYear{2019}
\setcopyright{acmcopyright}
\acmConference[COINS]{INTERNATIONAL CONFERENCE ON OMNI-LAYER
	INTELLIGENT SYSTEMS}{May 5--7, 2019}{Crete, Greece}
\acmBooktitle{INTERNATIONAL CONFERENCE ON OMNI-LAYER
	INTELLIGENT SYSTEMS (COINS), May 5--7, 2019, Crete, Greece}
\acmPrice{15.00}
\acmDOI{10.1145/3312614.3312646}
\acmISBN{978-1-4503-6640-3/19/05}

\begin{abstract}
With a rapid growth in the IT industry, Internet of Things (IoT) has gained a tremendous attention and become a central aspect of our environment. In IoT the things (devices) communicate and exchange the data without the act of human intervention. Such autonomy and proliferation of IoT ecosystem make the devices more vulnerable to attacks. In this paper, we propose a behavior monitor in IoT-Blockchain setup which can provide trust-confidence to outside networks. Behavior monitor extracts the activity of each device and analyzes the behavior using deep auto-encoders. In addition, we also incorporate Trusted Execution Technology (Intel SGX) in order to provide a secure execution environment for applications and data on blockchain. Finally, in evaluation we analyze three IoT devices data that is infected by mirai attack. The evaluation results demonstrate the ability of our proposed method in terms of accuracy and time required for detection. 
\end{abstract}

%
%

\ccsdesc[500]{Security and privacy}
\ccsdesc[300]{Security and privacy~Trusted computing}
\ccsdesc[300]{Security and privacy~Distributed systems security}
\ccsdesc[300]{Computer systems organization~peer-to-peer architecture}
%
\keywords{Blockchain, IoT, Security, Trust, Behavior, Deep Learning}

\maketitle

\section{Introduction}
Internet of Things (IoT) is rapidly increasing and currently involved in every field of our daily life. Industry leading experts argue that more than 50 billion devices will be interconnected by 2020 \cite{gartner}. These things are composed of web-enabled devices that use sensors, embedded processors and communication hardware in order to send/receive data from environments. As a result of such rich communication, it generates a large volume of data in turn to use for various dependent services.

Besides, IoT allows the evolution of several areas such as home to smart-home, health-care to smart-health-care, cities to smart-cities and many more. The key idea behind the IoT ecosystem is the diversity of things which results in a huge number of devices. Each device (physical or virtual), should be traceable and the generated content/information can be retrievable by other users irrespective of their locations \cite{lightweight}. Nevertheless, it is necessary that only authorized users can be able to access and make use of the system. Otherwise, it may become more prone to several security issues such as information leakage, data modification and identity theft. Furthermore, security issues remain a promising challenge in such a large scale adoption of IoT system because of the reasons: (1) Since most of communications between these devices are wireless which make the system more vulnerable to different attacks, i.e. eavesdropping, message tampering and \emph{mirai} attack \cite{mirai} etc. (2) Devices from different manufacturers have limited capacity in terms of processing, battery, and memory that do not allow to implement advance security solutions.

A number of solutions regarding security in IoT have been proposed that offer the mainstream security requirements i.e. Confidentiality, Integrity, Authentication or simply CIA \cite{cia}. However, due to heterogeneity and resource-constraint devices, existing solutions cannot fulfill the gap of security in such a huge expected IoT system. Furthermore, some of solutions are although efficient and secure but are often based on centralized mechanisms. For instance, PKI (Public Key Infrastructure) faces scalability issues in case of thousands number of nodes.

When it comes to decentralization, Block-chain (BC) technology has gained an overwhelming attention in regard to addressing security, auditibility, anonymity and centralization. Ethereum \cite{ethereum} a public blockchain was introduced in 2014 that deploys smart-contracts for BC users in order to write and execute code in a distributed way. Basically, BC is a decentralized ledger technology where each operation is recorded in the form of a transaction i.e. CRUD (create, read, update, delete) in context of IoT. Any unauthorized access to data or any operations on the previously stored data can, therefore, be detected. Moreover, smart contracts are used to enforce certain access control mechanisms on the stored data. A number of researchers put efforts to make BC technology suitable for IoT use cases \cite{dorri2017towards} \cite{platibart} \cite{iot} \cite{Ali2018} \cite{Roulin} \cite{sharing} \cite{pharma}.

\subsection*{Problem Statement and Contribution:}
As from various studies, it has been found that blockchain has become a promising technology to meet upcoming IoT security requirements \cite{Christidis2016}. Authors in \cite{authzone} proposed a mechanism of decentralized authentication in different zones of IoT systems. But the limitation to this approach is: there is no device-level trust that can prove any particular zone to external entities in case of supposing the communication to occur between different IoT use-cases.

The contribution of this paper is two-fold:
\begin{enumerate}
	\item Proposed a \emph{Behavior Monitor} in IoT-Blockchain infrastructure that can store IoT devices data and classifies behavior (normal or malicious) in order to prevent attacks.
	\item To incorporate Trusted Execution Environment on a local blockchain (i.e. Hyperledger) of each IoT-Zone that guarantees the integrity and confidentiality of application code and data.
\end{enumerate}

\subsection*{Outline}
Section \ref{bac} discussed background related to proposed framework. In Section \ref{rel} some past efforts in the integration of IoT and Blockchain are mentioned. Section \ref{arc} discusses the proposed architecture and all its components in detail. In Section \ref{eval} we demonstrate some initial results we have found while making experiments. Finally Section \ref{conc}  concludes the proposed work and discusses the future work.

\section{Background}\label{bac}
\subsection{Blockchain}
Blockchain technology was initially introduced and adopted by a notably known cryptocurrency, Bitcoin \cite{bitcoin}. BC is a decentralized ledger technology that relies on a peer-to-peer network. Each node in the BC network keeps an updated copy of ledger which can prevent from a single point of failure. 
In the past few years, the blockchain specifically functioning on cryptocurrencies \cite{bitcoin} \cite{hyperfabric}, i.e. to avoid double spending problem. However, recently numerous application areas have been explored where the blockchain can be deployed to create and keep digital records (transactions) in a distributed and secure way.

\begin{figure}[!h]
	\centering
	\includegraphics[width=\linewidth]{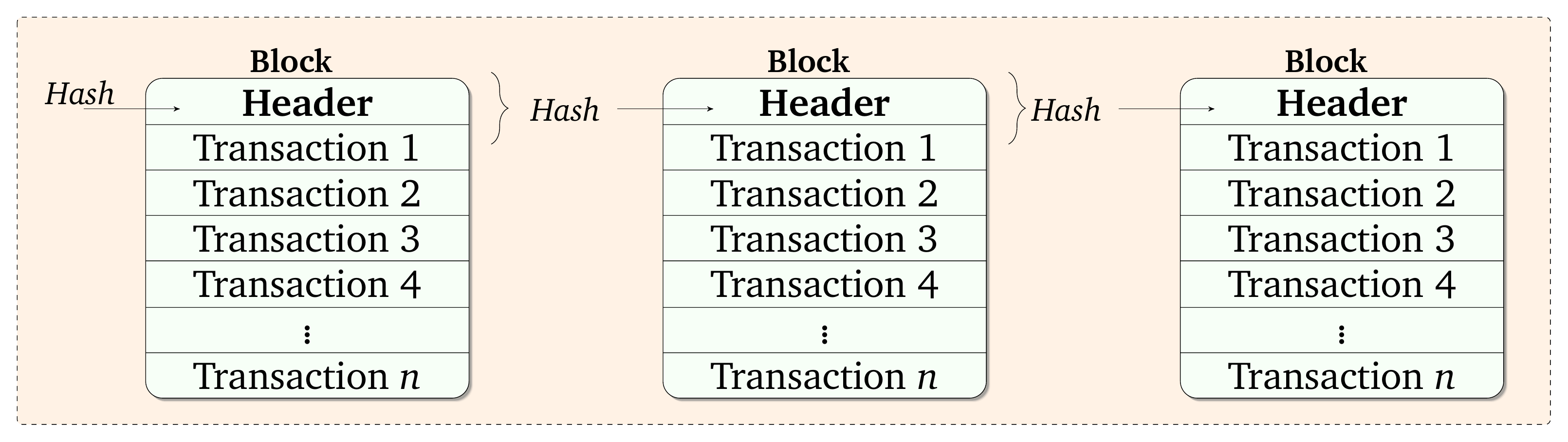}
	\caption{Linked Blocks}
	\label{bc}
\end{figure}

The ledger in BC is consists of blocks and each block comprises two parts. The first part represents the transaction or a fact (that need to be stored in a database), which can be of any kind, such as goods transaction, healthcare record, network traffic log etc. The second part includes the header information, i.e. hash of transactions, hash of previous hash, timestamp. Thus, storage in this way makes a linked chain of sequenced blocks as shown in Figure \ref{bc}. Moreover, if a new transaction needs to add to BC, it will first add to some block. Miners verify the block contain the transaction according to defined rules. After verification, all miners perform a consensus mechanism to validate the transactions. Finally, after successful validation the verified transaction will be added to BC ledger.   

\subsection{Blockchain and IoT Systems}
IoT devices produced an enormous volume of data, which must be stored and analyzed. For each IoT operation (create, update, delete, read), the data can be registered in the form of transactions in the BC-blocks. Identity information of IoT device can be registered in a block such as manufacturer information and current status where the device is in use. Smart-contracts are used to apply access control policies for IoT devices which means that any unauthorized access to a device can be therefore detected. There is no need of centralized mechanisms for storage, such as cloud, for IoT protection. Blockchain provides data authenticity and prevent from unauthorized access. BC can also enable a secure way of messaging between IoT devices. Exchange of messages can be treated like financial transactions flow in crypto-currencies e.g. Bitcoin \cite{bitcoin}. 
\subsection{Blockchain Security Solutions for IoT}
The decentralized and distributed nature of blockchain makes it a promising solution for IoT security. IoT with integration in blockchain enable a higher security level, which otherwise could not be achieved by any other technology or nearly impossible. Some of recent proposed solutions in terms of IoT security with blockchains are as follows:

In \cite{managing}, authors proposed a blockchain based solution for IoT device control and configurations using Ethereum. A unique key-pair (Public \& Private) is assigned to every device. The private key is stored inside device, while the public key is registered as a transaction in the blockchain. An IoT device can then be reached through Ethernet using its public key. Thus, it is concluded that the management of IoT devices via blockchain is possible.

A solution proposed in \cite{lee}, which make use of blockchain for secure firmware updates in IoT devices where traffic directly to the network server is replaced by local peers of the blockchain nodes. The manufacturer is supposed to store the hashes of updated versions of firmware on the blockchain which can be accessible to all the IoT nodes.

IoT devices related to medical are also subjected to the same security and privacy concerns. A medical IoT system must be attack resistant and reliable enough. User safety and privacy in this case is very crucial. A user must be protected from any malfunction caused by a security incident or faulty device. Blockchain can overcome the risk of device malfunction by immutable ledger of records. Nichol et al. \cite{medical} proposed the applicability of BC in order to ensure reliability in medical IoT devices. Upon a device is produced and deployed, a hash of unique ID along with the other relevant information such as manufacturer name, are stored in BC. Later, this data will be updated with patient history, doctor name and hospital information. The doctors and patients can be automatically notified about the device status, i.e. expiring battery, patient health irregularities. 
\subsection{Blockchain \& Trusted Execution Environment (TEEs)}
Trusted Execution Environments (TEEs) \cite{sgxonline}\cite{sgxonline1} have been utilized to strengthen security and performance in the blockchain protocol. TEEs provide confidentiality and integrity of the application code in a system, until and unless the CPU is not compromised by an intruder physically. TEEs also support remote attestation \cite{johnson2016intel}, that allows remote parties to verify the health of software with genuine TEE. 

Intel provided TEEs functionality in \emph{Software Guard Extension} (SGX) \cite{sgxonline}. SGX is a set of CPU instructions inside Intel's x86 processor design which can allow to create an isolated environment for the execution of selected pieces of code in protected areas called \emph{enclaves}. These enclaves are designed to run software in a trustworthy way, even on a system (host) where the operating system and memory are untrusted. There are three main functions of enclaves which are \emph{isolation}, \emph{sealing} and \emph{attestation}. A short description are as follows:
\begin{itemize}
	\item Isolation: Data and code inside the enclave memory are protected and cannot be read or altered by any external process.
	\item Sealing: Data supposed to send to host environment should be encrypted and authenticated with a \emph{seal} key.
	\item Attestation: Remote parties are allowed to verify an application enclave identity, credentials and other data.
\end{itemize}

\subsection{Threat Model}

The threat model for this research are as follows:
\begin{itemize}
	\item The proposed solution will work on permissioned blockchain with trusted execution technology enabled systems.
	\item Trusted execution environment (TEE) is considered as a root-of-trust so the TEE, CPU and hashing algorithm are considered trusted.
	\item Network between IoT devices to the behavior monitor system is considered secure. No assumptions are made about software running alongside the behavior monitor. There can be any number of malwares trying to exploit the transactions by IoT devices in the internal network.
\end{itemize}

\section{Related Work} \label{rel}
Currently, several researchers show interest in the integration of blockchain and IoT ecosystems. Very few of them are related to help IoT security requirements. This section outlines some of the past research efforts that intend to realize such integration particularly for security needs. 

Dorri et al.\cite{dorri2017towards} deploy blockchain based architecture for smart-home setting. Their approach consists of three different blockchain networks: a local-BC (private) for every use case, a share BC (private) and overlay BC (public). Although this solution solves the issue of identification, but still it has some limitations like: (1) For each operations it happened to create at least 8 communication links that can flood the network easily in case of high activity of IoT devices. (2) Local BC's are centralized and not distributed which is opposite to the main principal of BC - a decentralized technology. 

In \cite{fairaccess}, authors study existing solutions regarding access control systems and argue that these systems are not effective in the domain of IoT because of its expected growth from millions to billions of devices. In order to get rid of centralized mechanisms, this proposed solution implements capability and access control as a component in a blockchain infrastructure. The other components are: data management protocol, messaging service and data storage system. The messaging service ensures the exchange of access control message between two parties with defined roles. The messaging service then send the request to a data storage system, where it is stored as a block. Finally, the receiving entity fetches the message from the BC block using the messaging service. They further defined four roles, i.e. data source, data owner, requester and endorser. 

Hardjono et al.\cite{hardjono2016cloud} propose a privacy preserving mechanism called \emph{chainanchor} for authorizing IoT devices in the cloud network. It helps device-owner being rewarded when selling their device data to a service provider and ensure a privacy preserving communication between service provider and device-owner. But this approach is not suitable in most IoT use-cases, because the main objective of this approach is full anonymity and IoT devices sometime need device identification.

Patrick et al. \cite{authzone} propose a decentralized authentication scheme for IoT devices. In this approach they declare virtual zones such as smart-home zone, healthcare zone, for robust identification of smart-devices. Each zone has a group master who is responsible to create a groupID and communicate with blockchain. Each device or follower in a zone gets a ticket signed by their respective zone master. When a device or follower want to start a transaction, it will first send an association request signed by private-key to their respective zone master. Upon receiving the request, blockchain verifies the integrity with the follower public key. Then the follower ticket is verified using the master public key. If the ticket is valid, BC stores the association, i.e. groupID with followerID for further correspondence, otherwise discarded. However, the shortcoming of this approach is that there is no trust-level integrity measurement of each zone devices in order to prove it to the outside community.

To summarize, majority of all these current researches follows the same security mechanism provided in existing BC technologies i.e. Bitcoin \cite{bitcoin}, Ethereum \cite{ethereum} etc. However, there is no awareness towards device level trust that means to know about the device state (benign or malicious).
\begin{figure}[!th]
	\centering
	\includegraphics[width=\linewidth]{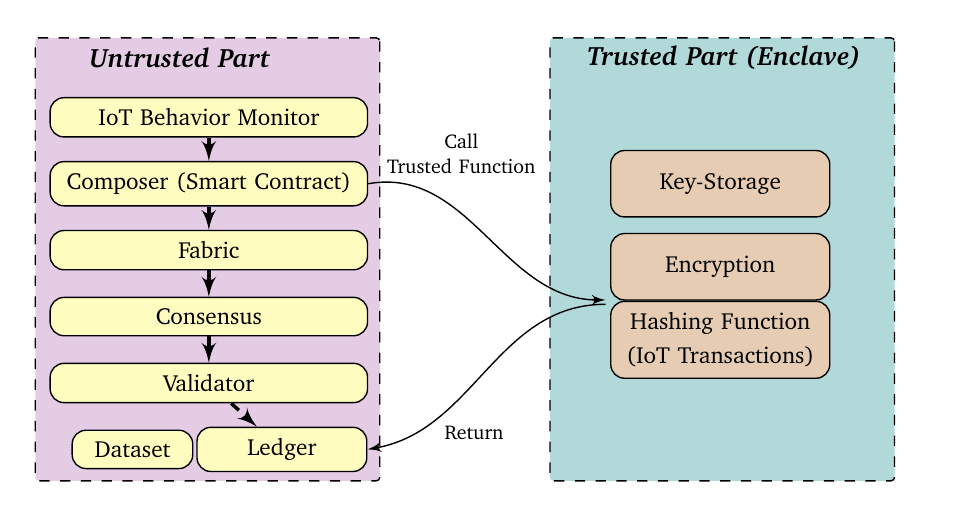}
	\caption{IoT secure behavior capturing and storage mechanism}
	\label{en}
\end{figure}

\section{Proposed Architecture} \label{arc}
\begin{figure*}[!th]
	\centering
	\includegraphics[width=0.67\textwidth]{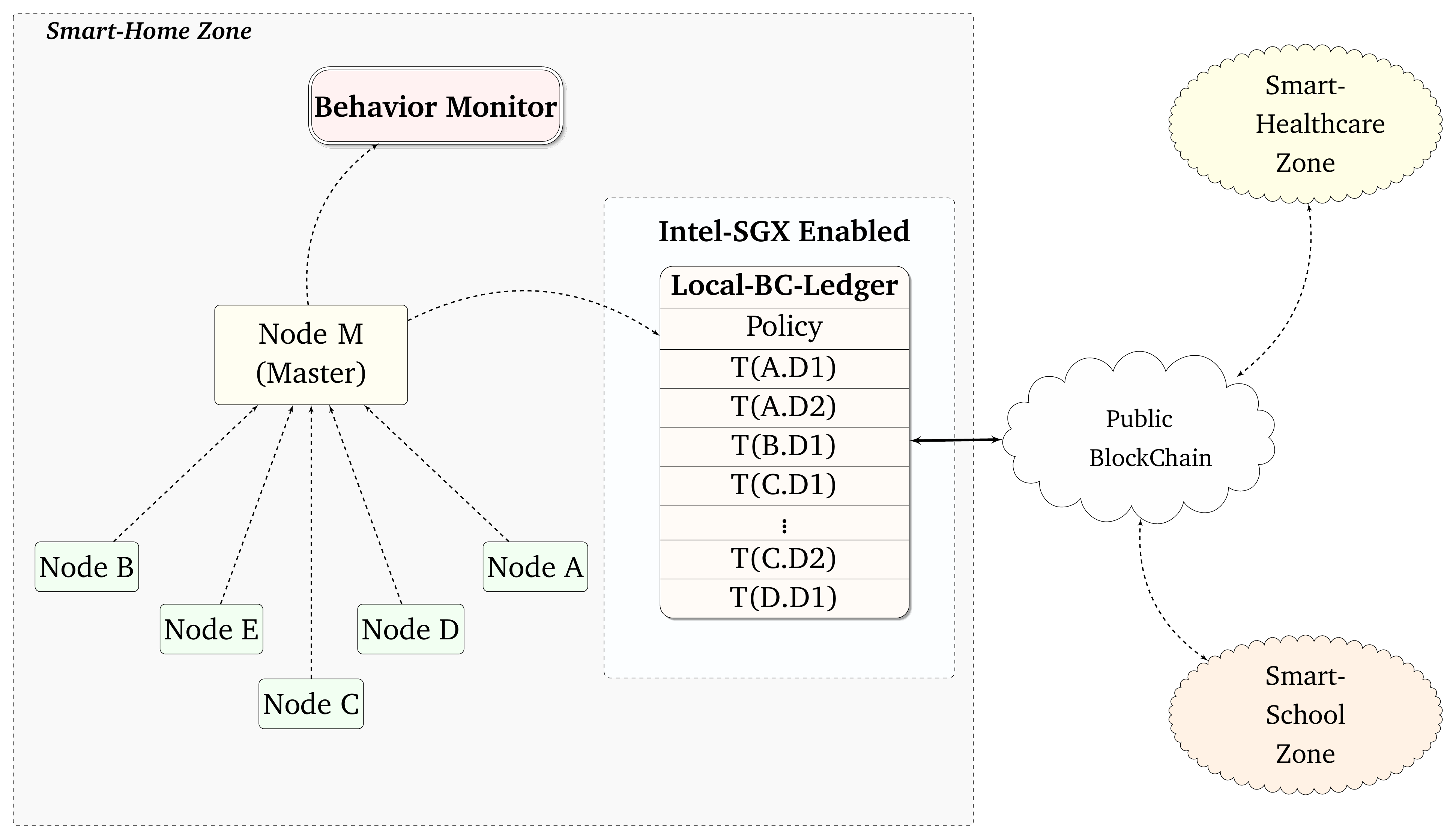}
	\caption{Proposed IoT Blockchain Architecture} \label{arch}
\end{figure*}

The main goal of the proposed architecture (cf. Figure \ref{arch}) is to add a layer of security for behavior monitoring of various IoT-zones in a blockchain setup. As discussed in \cite{authzone}, authors declare zones for different use-cases of IoT, however, they do not consider the devices itself in case of compromised behavior. Furthermore, there is no mechanism that can show the level-of-trust for each zone when an external entity needs to know before communication. In this research, we enhance the said scheme and add a behavior monitor on each zone. A separate local-BC is configured on each zone that is used to store the activity of each zone and provides the trust-level confidence to outside entities.
All kinds of communications between devices are considered as transactions and must be passed through the blockchain for validation. For example, if Node \emph{A} need to send message to Node \emph{B}, then \emph{A} must first send the message to blockchain. If BC validate and authenticate the message from \emph{A}, then \emph{B} is finally allowed to read the message.
The proposed architecture follows the same procedure of devices initialization and system functioning discussed in \cite{authzone}. For brevity, we discuss a brief description in the following section.
\subsection{Initialization \& System Functioning}
In the first phase of deployment, one device from each zone is designated as a \emph{Main or Master} node which can be considered as a certification authority (CA). Any node can be defined as a master, but in this case, we assigned to the node that is more resource capable and powerful. All the other nodes in each zone are known as \emph{follower}. Every Master node creates a \emph{groupID} and send a signed \emph{ticket} to each follower for identification. For the first transaction of any follower, it must require authentication. After that, an association of the follower and master are stored in the BC for future correspondence. For more detail about the initialization and system functioning we refer the reader to \cite{authzone}.

\subsection{Local Blockchain Setup}
A local blockchain is deployed on every zone and populated with the hashes of transactions generated from IoT devices. Hyperledger Fabric \cite{hyperfabric} is chosen as a local BC, we discussed the details of fabric with IoT in our previous work \cite{Ali2018}. 
For prototyping, we use the dataset \cite{dataset} of IoT traffic collected from various sensor communication. For each communication between devices or nodes, a transaction is created and stored in the local BC. Note that in most of the existing BC technologies, actual data of IoT devices are not stored in the BC due to processing overheads. 

In each zone a single device having more computational power than others, acts as a \emph{master} or \emph{main} node. Once the number of transactions reaches to a pre-define \emph{blocksize}, the master node creates a new block and append it to local BC. Afterwards, we incorporate \emph{Intel SGX} \cite{sgxonline} as a root-of-trust on top of BC in order to ensure that the execution of code and applications are trusted. As shown in Figure \ref{en}, the Intel SGX-enabled application is composed of trusted and untrusted part. For sensitive operations i.e. encryption, hashing, a trusted-function is called. The function returns, and the data inside the trusted part (enclave) remains in trusted memory and are not accessible to external entities. Moreover, implementing \emph{SGX} technology over blockchain allows the proposed scheme to:
\begin{itemize}
	\item Protect the applications running on BC and data protection that cannot be accessed by the execution host.
	\item Make sure that the application/data on BC is expected and correct.
	\item Protect end-to-end privacy of application result, which cannot allow others to inspect but the user.
	\item Provide a BC-based validation by verifying the applications inside \emph{enclave} is neither tampered nor interrupted by any node in BC.
	\item Make sure the application and execution results are valid, and not tampered or fabricated by any malicious node.
\end{itemize}


\subsection{Behavior Monitor}
The main goal of this research is to integrate our own custom behavior monitor that can classify the behavior of every device and compute a level-of-trust on each zone. As mentioned earlier, that all the nodes (followers) in a specific zone do their operations (read, write) via master/main node. The scheme in Figure \ref{arch} depicts our proposed approach with all the entities in detail.  Data or transactions from nodes is considered as a behavior of that particular node. Master node is a device that centrally processes all the incoming and outgoing transactions to and from a zone. 

Whenever a data is received by the master node from the follower node, the master node stores the data in behavior monitor and append the corresponding hash to the ledger in blockchain. A sequence-ID (SEQ-ID) is assigned to each transaction while storing in behavior monitor, and a Hash-ID (H-ID) is assigned to the corresponding hash in BC, for reference. Finally, a machine learning strategy is used to actively monitor the incoming data and classify them as normal or malicious. 

For analysis and detection of behavior we rely on deep auto-encoders \cite{hinton2006reducing} \cite{nauman2017deep} for IoT devices, which is trained from statistical correlation features extracted from benign data. The process of behavior detection and monitoring consists of the following stages. (1) Data collection (2) Feature extraction (3) Training model (4) Continuous Monitoring.
\subsubsection{Data Collection}
At this point, we refer to the dataset \cite{dataset} that has been collected from various sensors in IoT network. In real-time, in order to ensure that the training data is clean and not malicious, normal traffic from IoT devices are collected immediately after its installation to the network.  
\subsubsection{Feature Extraction} 
Whenever data from IoT device arrives, a behavioral snapshot of the protocols and host related to data are stored in our \emph{behavior monitor}. The snapshot contains different parameters, i.e. source IP, destination IP, MAC-address and port number, etc. We use the same set of features mentioned in the dataset for real time detection of malicious activities in IoT devices. For example, when a compromised node in a zone spoof an IP, then the features aggregated from the source-IP, destination-IP and MAC-Address will immediately mark as malicious because of unseen activity from the respective spoofs IP.

\begin{figure*}[!th]
	\begin{minipage}[b]{.44\textwidth}
		\includegraphics[width=\linewidth]{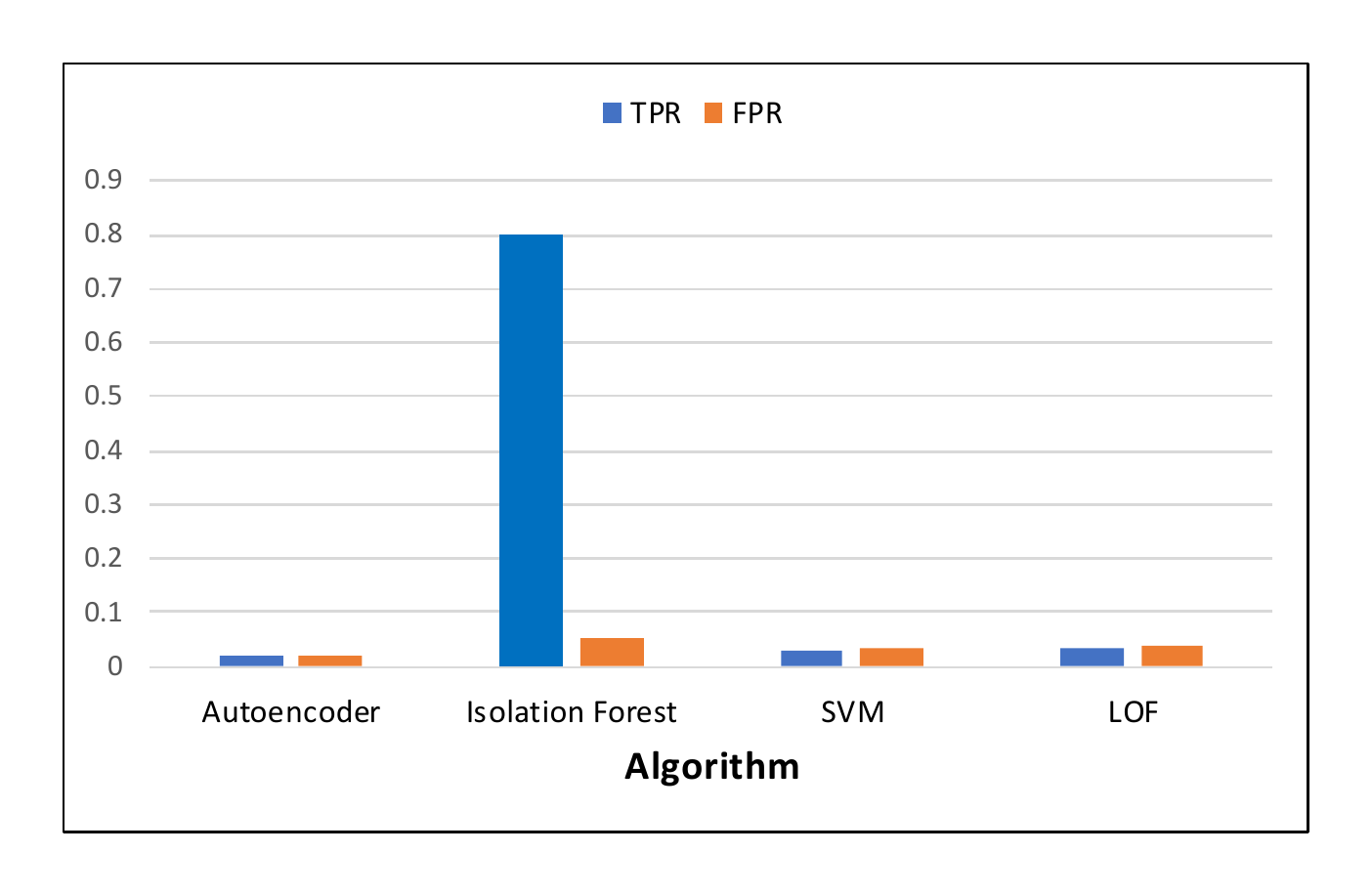}
		\caption{Algorithms' detection accuracy}\label{det}
	\end{minipage}\qquad
	\begin{minipage}[b]{.45\textwidth}
		\includegraphics[width=\linewidth]{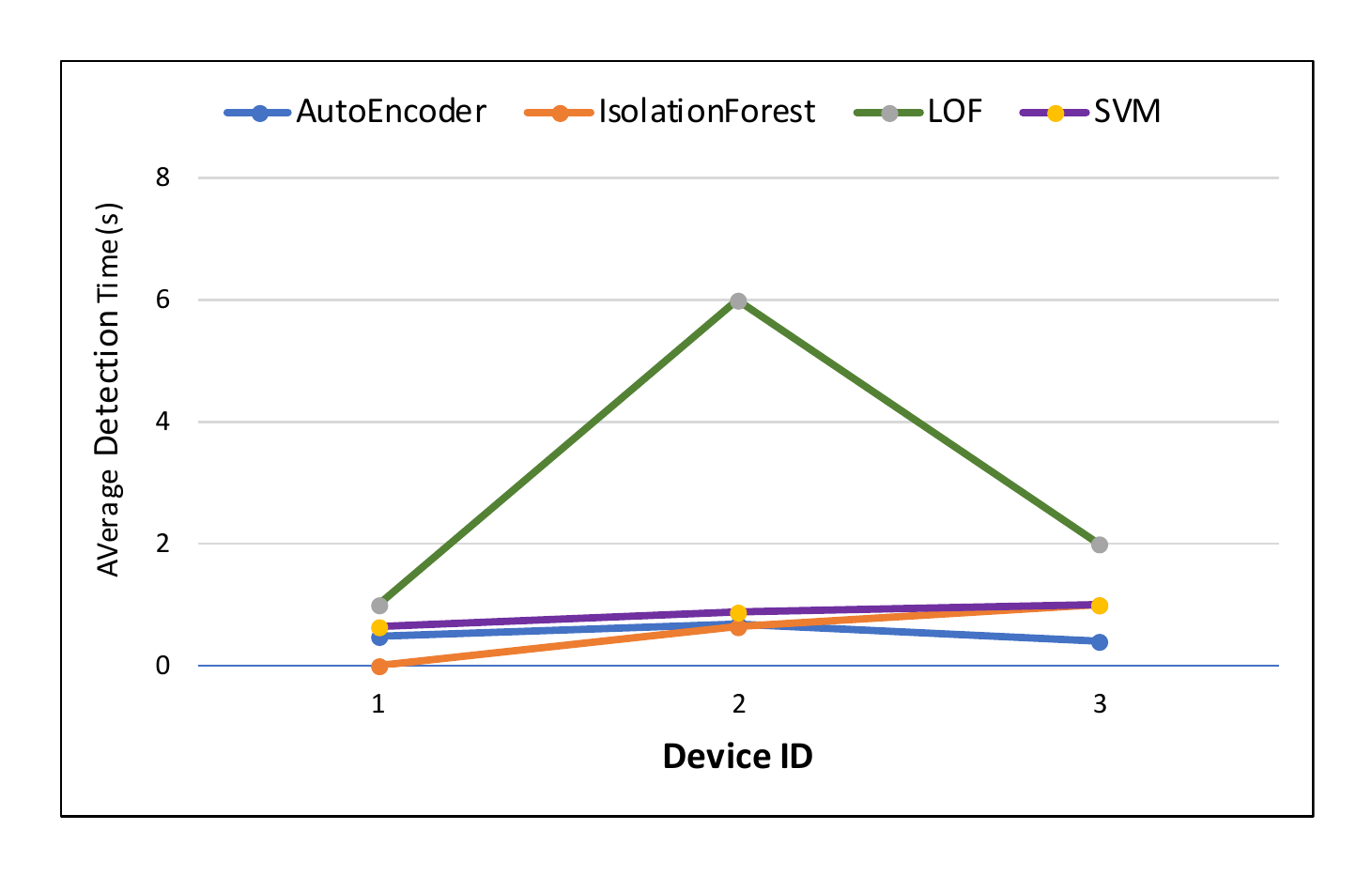}
		\caption{Algorithms' detection time (seconds)}\label{time}
	\end{minipage}
\end{figure*}

\subsubsection{Training Model}
As our baseline model for behavior detection, we use deep auto-encoders that can build and maintain a learning model on each zone of IoT use-cases. An auto-encoder is a type of neural network, which is trained to re-organize the data after some sort of compression. The compression makes sure that the model learns meaningful concepts and the correlation between different set of features.
For training purposes, we use two sets of data which consists of only benign (normal) data. The first dataset is a \emph{training dataset} $(T_{DS})$ which is used to train the auto-encoder by declaring input parameters such as \emph{learning rate} $(lr_{n}$, size of gradient descent step), and $epochs$ (number of iterations through $T_{DS}$). The second dataset $Opt_{DS}$ \emph{(Optimization Dataset)} is used to optimize the above hyper-parameters ($lr_{n}$ \& $epochs$) iteratively until the mean square error $(MSE)$ function between the input and output stop decreasing. This stopping prevents overfitting in $T_{DS}$ and helping in better detection results with future data. Later on, $(Opt_{DS})$ is used to segregate between normal and malicious activities and false positive rate (FPR).

After the model training and optimization is completed, threshold value $(th^v)$ is set by which an instance of data is considered malicious.
Empirically, it is calculated as the sum of the sample mean and std\_deviation of $MSE$ on $Opt_{DS}$ (see Equation).
\begin{displaymath}
th^v = \overline{MSE}_Opt_{DS} + s (MSE_Opt_{DS})
\end{displaymath}
\subsubsection{Continuous Monitoring}
Finally, the model is applied to continuously observe the data and to label each instance as a normal or malicious. Consequently, an alert against anomalous behavior can be issued in order to indicate the IoT device is malicious.

\section{Evaluation}\label{eval}
In our experiments, we use a real-time dataset available in \cite{dataset}, for realizing the framework. The dataset contains both the benign and malicious (attacked) data. The data we choose from the dataset belongs to three different devices, i.e. Ecobee-thermostat, Webcam and Security-camera. For training and optimization, we use \emph{tensorflow} and \emph{keras} libraries in python language. An auto-encoder make an input layer whose dimension is the same as the number of features in the dataset i.e. 115. 

After training we apply a famous attack (\emph{mirai}) to compute the detection time and accuracy of our model in comparison with other algorithms commonly used for anomaly detection. The same benign dataset is used to train three other algorithms: SVM (support vector machine), Isolation forest and LOF (\emph{Local Outlier Factor}). Our method shows 99.8\% results in terms of TPR (True positive Rate) and fewer FPR (False positive Rate). Furthermore, as evident in Figure \ref{det} SVM and LOF have almost similar TPR value and found much better than the isolation forest. Next, we evaluate the average detection time for each algorithm as depicted in Figure \ref{time}. The detection time of all the three devices in our case is lower than the others.


The deep auto-encoders dominate on all the selected devices in terms of TPR, FPR and detection time. This is because of the ability in auto-encoders to learn approximate complex functions and non-linear structure mapping \cite{detection}. 


Moreover, as shown in Figure \ref{time}, our technique required much less time than the other algorithms which is approximately 170$\pm$220ms (milliseconds) to detect the attacks. This means that the launch attack could be detected or alerted in less than a second and thus considers as a  substantial reduction in a typical time required for DDOS attacks \cite{DDOS}.
\section{Conclusion and Future Work}\label{conc}
In this research, we analyze device level trust in IoT-Blockchain Infrastructure. A smart-home setting is used as a use-case for realizing the proposed idea. A Local Blockchain on each zone is deployed that can store every traffic coming from their follower (nodes) in the form of transactions. Behavior Monitor is defined and configured on the Main/Master node of each zone, which can capture and analyze the activity of IoT devices. We apply a deep learning strategy (auto-encoders) on the behavior monitor to classify the device and make a level-of-trust. Furthermore, we incorporate Intel SGX technology as a root-of-trust over the blockchain to provide security for sensitive code and applications. Finally, the evaluation of our study shows its ability to mitigate the mainstream security requirements and resilience to attacks.

This research is our preliminary step towards classification of devices in  IoT-Blockchain framework. Our future plan is to investigate a comparative study of other machine learning approaches for better performances and accuracy. Another goal would be to apply the framework to other application areas of IoT domain and analyze the outcomes. Finally, we will provide a full implementation on various IoT devices datasets and technical details about the Intel SGX implementation.

%
\bibliographystyle{ACM-Reference-Format}
\bibliography{sample-base}

%

\end{document}